\documentstyle{article}

\begin{document}
\title{Second order contributions to the absorption of massive particles}
\author{Pedro Sancho \\ GPV de Valladolid \\ Centro Zonal en
Castilla y Le\'on \\ Ori\'on 1, 47014, Valladolid, Spain}
\date{}
\maketitle
\begin{abstract}
Recently, in analogy with multiphoton ionization, it has been
suggested that multiparticle ionization can also be induced by
massive systems. We explore in this paper the possibility that
multiparticle absorption processes can also take place for massive
particles. To study it we consider, in a perturbative way, a model
of absorption which illustrates the analogies with Glauber's
scheme for photons and previous analysis on matter-waves
coherence. A major advantage of this approach is that the
dependence of the absorption rates on the wavefunction of the
incident system can be analyzed in an explicit way. The
calculations confirm the form of the second order (two-particle)
contributions.
\end{abstract}
\vspace{7mm} PACS: 03.65.-w; 42.50.Hz; 03.75.Be


\section{Introduction}

Multiphoton processes have been an active area of research during
the last decades \cite{mul}. Recently, it has been suggested
\cite{San} that similar events can occur in the case of massive
particles. An example where this possibility can be easily
visualized was presented in Ref. \cite{San}. It is the process of
ionization. One incident particle can ionize the atoms or
molecules of a piece of matter or a gas. This is the usual way we
think about ionization. However, there can be other ways. For
instance, if two particles arrive on the piece of matter or gas
one of them can excite the atoms, with the second one producing
the definitive ionization. The similarity with multiphoton
ionization is clear. This type of process can be important because
of the recent arrival of atom lasers and other atomic systems
showing high degrees of coherence, in close analogy with light
lasers.

Massive and massless particles can be absorbed by matter in various physical
processes.
Absorption has been extensively studied in the case of photons.
By analogy with the considerations about ionization we can presume that
multiparticle absorption
could also take place in the massive case. We shall analyze in this paper this
possibility.
In Ref. \cite{San} it was shown that multiparticle processes would imprint a
clear mark that would distinguish them from single particle events: the
probability of a multiple process is not proportional to the squared modulus of
the
wavefunction, as it is the case for single events. In this paper we want to
confirm that result in the case of absorption, but using a more physical
approach to the problem. Instead of
relying on the formal approach of Ref. \cite{San} where only the probability of
detection was considered, we shall introduce here a model where the explicit
form of the interaction is presented. Our analysis will be developed in analogy
with Glauber's scheme. Many years ago, Glauber \cite{Gla} introduced a
mathematical description of photon absorption which has been extensively
verified and provides a solid basis for the understanding of the detection of
light beams in photomultipliers and other measurement devices.

We extend in this paper Glauber's scheme to the case of massive
particles. This is not the first time that the ideas of Glauber
have been applied in the framework of matter waves. In Ref.
\cite{Gol} different types of atomic coherence were considered in
connection with different types of atomic measurements. Even in
Ref. \cite{Pat} the mutual coherence of optical and matter waves
have been studied. In particular, these authors considered three
types of measurements, fluorescence, nonresonant imaging and
ionization. However, these authors did not consider the case
explicitly studied here, the absorption of massive particles.
Another important difference between the approach of Refs.
\cite{Gol,Pat} and our analysis is that the first one was mainly
concerned with the properties of matter waves, whereas the second
one is mainly aimed at the study of the dependence of the
absorption.

We emphasize the last point. Our approach to multiparticle
absorption is focused on the dependence of the absorption
probabilities on the wavefunction of the incident beam. This point
of view is novel in the case of incident massive beams. Previous
approaches to the problem rely on either (i) a classical treatment
of the intensity of the incident beam, or (ii) a quantum
description in the framework of scattering theory. Imaginary
absorption coefficients or optical-type potentials \cite{GG} are
introduced to simulate the interaction. However, in the first case
no information can be obtained about the relationship between the
absorption and the form of the wavefunction, and in the second one
it is very difficult to extract.

The plan of the paper is as follows. In Sect. 2 we introduce our
model of detection of massive particles by extension of Glauber's
scheme and in analogy with previous studies of matter-waves coherence. Sections
3 and 4 are devoted, respectively, to the
evaluation of the first and second order of the perturbative
expansion of the model. At first order we obtain the transition
rate for incident one-particle beams, which is proportional to the
squared modulus of the wavefunction. At second order, which
describes two-particle absorptions, we find a different type of
dependence on the wavefunction. Finally, in the last section we
discuss the main results of the paper.

\section{A model for the absorption of massive particles}

In this section we present our model for the absorption of massive particles. As
signaled before, the model is an extension of Glauber's theory for the
absorption of photons to the case of particles ruled by Schr\"{o}dinger's
equation. Then our starting point will be a short review of the Glauber
approach.

\subsection{Glauber's scheme}

The theory of absorption of photons was originally introduced by Glauber
\cite{Gla}. It is a very general description and with minor modifications can be
applied to a large number of processes. For instance, it has been extensively
used for the description of detection devices such as photomultiplier tubes,
photographic plates or bolometers. In this scheme, the absorption is
mathematically represented by the action of the annihilation operator of the
photon under consideration.

The interaction between the incident beam and the device is also
specified. For instance, in the case of the photomultiplier it
is given by the electric dipole interaction Hamiltonian:
\begin{equation}
\hat{H} _{int}^{pho} =-\bf{d} . \hat{\bf E} ({\bf r_{pho}})
\end{equation}
where $\bf{d}$ is the atomic electric dipole and $\hat{ \bf{E}
} $ is the electric field operator at the location ${\bf r_{pho}} $
of the photomultiplier. As it is well-known the electric field operator is
decomposed in the form $\hat{\bf E}= \hat{\bf E} ^+ + \hat{\bf E} ^- $, where
each
part can be expressed in terms of the creation and annihilation operators of the
photons.

Using the interaction Hamiltonian and the expression for the
incident beam in Fock's space it is possible to calculate the
transition rate for the absorption.

\subsection{The model}

We consider massive particles that interact with a system or
medium, which has the capacity of absorbing them. Mathematically,
the absorption of photons is represented by annihilation
operators. Similarly, we shall describe the absorption of massive
particles by the action of the annihilation operators of the
particles. The annihilation operators describe absorption
processes but without any spatial localization. However, in
general, we consider absorptions well-localized in space. For
instance, we can consider absorbing mediums of small size. The
description of well-localized absorption processes can be achieved
using $\hat{\psi}^+_{\Omega }$ and $\hat{\psi}_{\Omega}$, the
raising and lowering field operators (the last one, also named
Schr\"{o}dinger's field operator): the $\hat{\psi}^+ _{\Omega
}({\bf r}, t)$ ($\hat{\psi} _{\Omega }({\bf r}, t)$) operator
represents one that takes the state of the system to one with one
more (less) particle located at ${\bf r}$ at instant $t$ (with
spin $\Omega $) \cite{LL}. As it is well-known the
$\hat{\psi}_{\Omega } ({\bf r},t)$  operator can be written in the
form:
\begin{equation}
\hat{\psi }_{\Omega } ({\bf r},t) = \int d^3 {\bf q} \psi _{{\bf q} \Omega}
({\bf r})
\hat{a}_{{\bf q} \Omega} (t)
\end{equation}
where $\psi _{{\bf q} \Omega } ({\bf r})$ is a complete set of orthonormal
stationary wave functions, labelled by the index ${\bf q}$ (if the
index is not a continuous one the integral must be replaced by
a sum). Note that the time dependence is carried by the
annihilation operator. The most common choice for this set is
given by plane waves of momentum ${\bf p}$:
\begin{equation}
\hat{\psi }_{\Omega } ({\bf r},t) = \frac{1}{(2 \pi \hbar)^{3/2}}   \int
d^3 {\bf p} exp(i{\bf p}. {\bf r}/\hbar ) \hat{a} _{{\bf p} \Omega} exp(-
iEt/\hbar )
\end{equation}
with $\hat{a} _{{\bf p} \Omega }=\hat{a} _{{\bf p} \Omega }(t=0) $ and $ E={\bf
p}^2/2m$ the energy.

The raising and lowering operators act on the Fock space of the
incident particles, which is constructed in the usual way, $|0>$,
$|1>$....$|n>$, with $n$ the number of particles in the beam.

The absorbing medium is made of atoms and, in accordance, we assume that it must
obey the
laws of quantum mechanics. The state of the medium previous to the interaction
will be represented by $|M>$. If after the interaction one incident particle is
absorbed by the medium, its state will be $|M_1>$. Finally, if in the incident
beam there are $N$ particles and all of them are absorbed after the interaction
the state will be $|M_N>$. We can think of $|M>$, $|M_1>$ and $|M_N>$ as
collective states of the medium. Note that they can be but do not need to be described in the
Fock space of the medium (for instance, the description of the atoms of the
photomultiplier is done in the first quantization formalism, whereas that of the
photons is in the second one \cite{Lou}).

We must now introduce the interaction between the particles and
the medium. The interaction results in the absorption of a
particle at the point where the medium is placed (we assume by
simplicity, a very small size medium). Then the interaction must
be expressed in terms of the lowering operator, which, as
discussed before, represents the absorption at definite points.
The particular form in which the lowering and raising operators
will be present in the interaction Hamiltonian will be dictated by
analogy with the massless case. In that case it is the product of
the field operator by another operator representing properties of
the absorbing atoms. Similarly, we propose an interaction
Hamiltonian which is the product of the lowering and raising
operators by another operator related to the absorbing medium.
Therefore, we propose the following interaction Hamiltonian (in
the tensorial product of both spaces)
\begin{eqnarray}
\hat{H} _I ({\bf Q},t) = \alpha \hat{{\cal M}} (t) \otimes
\hat{\psi}_{\Omega} ({\bf Q},t) + h. c. = \nonumber \\
\alpha \hat{{\cal M}} \otimes \hat{\psi}_{\Omega} ({\bf Q},t) +
\alpha ^* \hat{{\cal M}}^+ (t) \otimes \hat{\psi}_{\Omega}^+ ({\bf Q},t)
\label{eq:uno}
\end{eqnarray}
where $\alpha $ is a constant depending on the type of particle
and medium considered, which gives a phenomenological measure of
the strength of the interaction between particle and absorbing system. On the
other hand $\hat{\cal M}$ is the operator representing the physical reaction of
the
medium to the incident particles (in the case of photons it would correspond to
the atomic electric dipole). With this choice we are adopting a
phenomenological point of view: we do not care about the precise
physical mechanism that produces the interaction and absorption, but only on the
fact that $\hat{\cal M}$ is responsible (through these complex mechanisms) for
the process.

The only definitive justification of this choice for $\hat{H}_I$ would be done
by the experimental
verification of its physical consequences, which will be derived in next
section. This is a
clear advantage of the model, it is a testable one.

In the above expression ${\bf Q}$ is a parameter that indicates
the position at which the medium is placed. We introduce this
notation of ${\bf Q}$ as a parameter instead of the position
variable ${\bf r}$ to remark that $\hat{H} _I ({\bf Q},t)$ is not
a function of position but only depends on the value of the field
operator at ${\bf Q}$ ($ \hat{\psi} ({\bf Q},t)$). When we place
the medium at another position ${\bf Q_*}$ we must use a different
$\hat{H} _I ({\bf Q_*},t)$. For every value of ${\bf Q}$ we have a
different problem (see, for instance, Ref. \cite{LL} for the use
of ${\bf Q }$ as a parameter). The introduction of the position
variable as a parameter is a necessary consistency condition for
the cases where the problem is completely described in the second
quantization formalism (when the $|M>$ states associated with the
medium belong to Fock's space) because the Hamiltonian must be
position independent in the second quantization formalism.

The similarity between our approach and that presented in Refs. \cite{Gol,Pat}
is clear. In both cases, the interaction Hamiltonian is expressed in terms of
raising and lowering operators.

Before the interaction between the beam and medium
the state of the complete system is given by the product $|n>
|M> $. When both systems interact the state becomes an entangled
one of the type $|n,M>$. This entanglement describes the
correlation between both systems.

The Schr\"{o}dinger equation ruling the evolution of the complete system is:
\begin{equation}
i\hbar \frac{\partial}{\partial t} | n,M> =(\hat{H} _o
+ \hat{H}_I ({\bf Q},t)) | n,M>
\label{eq:dos}
\end{equation}
with $\hat{H} _o$ the free Hamiltonian of the complete system.

The similarities between our model and the optical scheme are
evident. One important difference between them must be
remarked now. Instead of the electric field operator we have in
the interaction Hamiltonian the raising and lowering operators of the
incident beam. The last two operators are expressed in terms of the
solutions of the Schr\"{o}dinger equation of the system, which
rules the behaviour of massive particles. Photons cannot be
encompassed in this category, because they do not obey
Schr\"{o}dinger's equation.

\section{Detection probabilities}

Using Eq. (\ref{eq:dos}) we can calculate the transition rate
between different states at a given position, i. e., the
transition rate at ${\bf Q}$ from the initial state $|n>
|M>$ to the state $|n-N> |M_N>$  with $N$ an integer number.
The transition rate is given by \cite{Lou}
\begin{equation}
w({\bf Q},t)=\frac{d}{dt} | <n-N|<M_N|\hat{U}({\bf Q},t)|M>|n>
|^2
\label{eq:cuat}
\end{equation}
where $\hat{U}({\bf Q},t)$ is the evolution operator of
(\ref{eq:dos}).

This equation must be computed in a perturbative way. The use of
perturbation theory in this problem can be easily justified. The interaction
between the atoms and molecules of the medium and the incident
particles are of the electromagnetic type, and it is well-known
that electromagnetic interactions can be described by
perturbation theory.

We shall evaluate the first and second order of the perturbative
development.

\subsection{First order}

To first order of perturbation theory Eq. (\ref{eq:cuat}) becomes \cite{Lou}:
\begin{equation}
w^{(1)}({\bf Q},t)=\frac{2\pi }{\hbar ^2} | <n-N|<M_N| \hat{H}_I
({\bf Q})|M>|n> |^2
\label{eq:bele}
\end{equation}
(in this expression, just as in the rest of transition
rates below, we also have a Dirac's delta between the energy of
the initial and final states, which is not explicitly included
in order to simplify the notation).

Taking into account the dependence of $\hat{H}_I ({\bf Q},t)$ on
the raising and lowering operators the above matrix element
is only different from zero for the lowering one and $N=1$. We consider as the
initial state of the particle $|n>=|1_{f \xi}>$ with
\begin{equation}
|1_{f \xi }>=\int d^3 {\bf b} f({\bf b}) \hat{a}_{{\bf b} \xi}^+ |0>
\end{equation}
where $f({\bf b})$ is the momentum distribution of the particle, obeying the
normalization condition $\int d^3 {\bf b} |f({\bf b})|^2=1$. In the first
quantization formalism this particle is represented by the wavefunction
\begin{equation}
\psi _{f \xi}({\bf Q})=\int d^3 {\bf b} f({\bf b}) \psi _{{\bf b} \xi} ({\bf Q})
\end{equation}
With that state Equation (\ref{eq:bele}) becomes $ w^{(1)}({\bf
Q},t)=2\pi | \Upsilon ({\bf Q},t) |^2 /\hbar ^2 $ with
\begin{eqnarray}
\Upsilon ({\bf Q},t) = \alpha <M_1|
\hat{\cal M}|M> \times \nonumber \\
\int d ^3 {\bf q} \int d^3 {\bf b} f({\bf b}) \psi _{{\bf q}
\Omega } ({\bf Q}) <0| \hat{a}_{{\bf q} \Omega } \hat{a}_{{\bf b}
\xi}^+ |0>
\end{eqnarray}
with $\hat{a}_{{\bf b} \xi}^+$ the creation operator at $t=0$.

Using the well-known (anti)commutation relations $[\hat{a}_{{\bf b}\xi},
\hat{a}_{{\bf q} \Omega}^+]_{\mp} =\delta _{\xi \Omega} \delta ^3 ({\bf b}-{\bf
q})$, where the upper sign is valid for bosons and the lower one for fermions,
we have $<0|\hat{a}_{{\bf b}\xi} \hat{a}_{{\bf q} \Omega}^+|0>= \delta _{\xi
\Omega} \delta ^3 ({\bf b}-{\bf q})$ (we assume the vacuum state to be
normalized $<0|0>=1$). Finally we obtain
\begin{equation}
w^{(1)}({\bf Q},t)=\beta \delta _{\xi \Omega }|\psi _{f \xi} ({\bf Q}) |^2
\label{eq:anph}
\end{equation}
with
\begin{equation}
\beta = \frac{2\pi }{\hbar ^2} |\alpha <M_1|\hat{\cal M}|M>|^2
\label{eq:pipi}
\end{equation}
Then the absorption probability (which is easily obtained from the
transition rate by integrating on time $w^{(1)}$) derived from
the model is proportional to the usual Born's distribution, which
establishes the probability to find a particle described by the wave function $
\psi  ({\bf Q}) $ at a point ${\bf Q}$ to be $|\psi ({\bf Q}) |^2$. In addition,
we have the coefficient $\beta $. A similar factor is present in quantum optics
where $w$ is the product of the efficiency factor and $I$ with $I$ the
quantum intensity of the field. Therefore, $\beta $ is to be identified with an
efficiency factor. It depends on $\alpha$, a function of the medium and the type
of particles considered, and on $<M_1|\hat{\cal M}|M>$ the matrix element of the
operator $\hat{\cal M}$. The efficiency factor incorporates into the theory the
possibility of interaction processes between medium and particle that do not lead to absorption. This is one of the possible channels of the
particle-medium interaction.

\subsection{Second order}

We consider now the second order of the theory. Physically, it
corresponds to incident beams with two particles (or
two different incident beams).

To second order the transition rate is $ w^{(2)}({\bf Q},t)=2\pi | \Xi ({\bf Q},t) |^2 / \hbar ^2$ \cite{Lou}, with
\begin{eqnarray}
\Xi ({\bf Q},t) = <n-z|<M_z|
\hat{H}_I ({\bf Q})|M>|n> + \label{eq:ulti}  \\
\sum _i (E_i)^{-1} <n-z|<M_z| \hat{H}_I ({\bf
Q})|M_{z-1}(I_i)>|(n-(z-1))(I_i)> \times \nonumber \\
 <(n-(z-1))(I_i)|<M_{z-1}(I_i)|
\hat{H}_I ({\bf Q})|M>|n>   \nonumber
\end{eqnarray}
and $z$ an integer.
The sum on $i$ refers to all the intermediate states of the
incident beam and medium. The energy $E_i$ is the difference
between the total (particle plus medium) initial energy and
that of the intermediate state $E_i=E_{po}+E_{Mo}-E_{pi}-E_{Mi}=
E_{po}-E_{pi}-E_{Mi} $ because we have taken
the energy of the detector in the initial state as the origin of
energies, i. e., $E_{Mo}=0$. $I_i$ in the intermediate states
refers to the two possible ways of absorbing the two particles.

We take now as initial incident state
\begin{equation}
|2>=\int d^3 {\bf b} f({\bf b}) \int d^3 {\bf d} g({\bf d}) \hat{a}_{{\bf b}
\xi}^+ \hat{a}_{{\bf d} \mu}^+ |0>
\end{equation}
The final state is $|n-z>=|n-2>=|0>$. The first term in the r. h. s.
of Eq. (\ref{eq:ulti}) is zero because of the presence of only one
lowering operator in the interaction Hamiltonian. If we assume that
the intermediate states $|M_{z-1}(I_i)> $ of the medium are
non-degenerate there are only two contributions to the second term
in the r. h. s. of Eq. (\ref{eq:ulti}). In the first contribution
the order of absorption is first $|1_{f \xi}>$ and later $|1_{g
\mu}>$, whereas in the second contribution the order is the opposite
one. The intermediate states of the beam and medium are,
respectively, $|1_{g \mu}>$ and $|M_{1}(f \xi )>$ and $|1_{f \xi}>$
and $|M_{1}(g \mu)>$.

Using the explicit expressions of the states and field
operators and applying repeatedly the (anti)commutation relations
we obtain the following relations (${\bf q}$ and ${\bf k}$ are the
integration variables in the field operator)
\begin{eqnarray}
<0|\hat{a}_{{\bf q} \Omega }\hat{a}_{{\bf u} \eta }^+|0> <0|\hat{a}_{{\bf u}
\eta }  \hat{a}_{{\bf k} \Omega } \hat{a}_{{\bf b} \xi }^+ \hat{a}_{{\bf d} \mu
} ^+|0> =  \label{eq:2un} \\
\delta _{\Omega \eta} \delta ^3 ({\bf q}-{\bf u}) (\delta _{\mu \eta} \delta
_{\Omega \xi} \delta ^3({\bf d}-{\bf u}) \delta ^3({\bf k}-{\bf b}) \pm  \delta
_{\xi \eta} \delta _{\Omega \mu} \delta ^3({\bf b}-{\bf u}) \delta ^3({\bf k}-
{\bf d}) ) \nonumber
\end{eqnarray}
Replacing ${\bf u}$ by ${\bf b}$ and ${\bf d}$ we can evaluate the transition
rate. After a simple but lengthy calculation we obtain:
\begin{equation}
w^{(2)}({\bf Q},t)=\frac{2\pi }{\hbar ^2} |\alpha |^4 |W({\bf
d},\mu)+W({\bf b}, \xi)|^2 \label{eq:2tr}
\end{equation}
with
\begin{eqnarray}
W({\bf d},\mu ) =<M_2|\hat{\cal M}|M_1({\bf d},\mu)>< M_1({\bf
d},\mu)|\hat{\cal M}|M>
\times   \\
(E_i({\bf d},\mu))^{-1} \delta _{\xi \Omega} \psi _{f \Omega}({\bf
Q})( \delta _{\xi \mu}  \delta _{\xi \Omega } <f|g> \psi _{f
\Omega } ({\bf Q}) \pm \delta _{\Omega \mu} \psi _{g \Omega } ({\bf Q}) )
\nonumber
\end{eqnarray}
and
\begin{eqnarray}
W({\bf b},\xi) =<M_2|\hat{\cal M}|M_1({\bf b},\xi)>< M_1({\bf
b},\xi)|\hat{\cal M}|M>
\times   \\
(E_i({\bf b},\xi))^{-1} \delta _{\mu \Omega} \psi _{g
\Omega}({\bf Q})(\pm \delta _{\Omega \mu}  \delta _{\xi \mu }
<g|f> \psi _{g \Omega } ({\bf Q}) + \delta _{\xi \Omega } \psi _{f \Omega }
({\bf Q}) )  \nonumber
\end{eqnarray}
with $E_i({\bf b},\xi)= E_p({\bf b},\xi)- E_{M_1({\bf b},\xi)} $, $E_i({\bf
d},\mu)= E_p({\bf d},\mu)- E_{M_1({\bf d},\mu)}$ and $<f|g>=\int d^3 {\bf
b}f^*({\bf b})g({\bf b}) $.

We shall consider two extreme cases where the implications of the new effects
can be seen more clearly.
The first situation occurs when the particles have no common modes, $<f|g>=0$
and they are in the same spin state, $\xi =\mu =\Omega $. Then the transition
rate is:
\begin{equation}
w^2 ({\bf Q}) \sim |\psi _{f \Omega} ({\bf Q})|^2 |\psi _{g
\Omega}  ({\bf Q})|^2 \label{eq:aster}
\end{equation}
The second case is when both incident particles are in the same state, $f=g$ and
$\xi =\mu =\Omega $ (note that this situation only refers to bosons, because of
the impossibility of preparing two indistinguishable fermions in the same
state). We obtain
\begin{equation}
w^2({\bf Q}) \sim |\psi _{f \Omega } ({\bf Q})|^4
\end{equation}
We see that the second order theory contributes to the
transition rate in the form $|\psi _{f \Omega}|^2 |\psi _{g \Omega}|^2$, or
for bosons in the same state as $|\psi _{f \Omega}|^4 $. This dependence on the
wavefunction differs from that of a single incident particle, which
is given by $|\psi _{f \Omega }|^2$.

The dependence of the absorption rate on the wavefunction is
reminiscent of that of photons on the quantum intensity $I$. In
this case the transition rate is proportional to $I$ for single
absorptions and to $I^2$ for double absorptions.

We also see that the multiabsorption rate for massive particles
differs for bosons and fermions. Because of the double sign in
all the above expressions the two terms add for bosons, whereas they
must be subtracted for fermions. In particular, when the states
$f$ and $g$ are close, the transition rate tends to the form
$|\psi _{f \Omega}|^4 $ for bosons and to $0$ for fermions.

\section{Discussion}

We have suggested in this paper the possibility of multiabsorption
processes in the case of incident massive particle beams. We have
presented a simple model to evaluate the probabilities of
absorption. We have corroborated the results obtained in \cite{San}
showing the existence of second order corrections, associated with
multiparticle processes, to the usual expressions for one-particle
processes.

The model presented here is simply the extension of Glauber's scheme for the
absorption of photons to the case of massive particles. With this generalization
we obtain a unified view for the absorption of both types of particles.
The model also has notorious similarities with the studies on matter-waves
coherence in Refs. \cite{Gol,Pat}. However, these authors did not consider
explicitly the case of massive particles absorption. Moreover, they were more
interested in coherence properties than in the dependence of the absorption rate
on the form of the incident wavefunction.

The model gives a more clear picture of the physical processes
involved than the formal approach to the problem given in Ref.
\cite{San}. For instance, we can relate the parameters
characterizing the strength of single and double absorptions (the
equivalent to parameters $\alpha _1$ and $\alpha _2$ in \cite{San}
introduced there only in a phenomenological way) with the physical
variables of the problem ($\hat{\cal M}$, $\alpha $, states of the
medium and particle...). A more detailed model than the one
introduced here would allow, in principle, for an explicit
calculation of the parameters.

Our approach to the absorption problem is fully quantum and
provides the possibility of analyzing in a simple way the
dependence of the absorption rates on the wavefunction.

The model can be tested experimentally. Sending one-particle beams
in different states towards the same absorbing medium (for
instance, placing the medium at different positions after an
interferometric arrangement, through which the incident beam has
previously passed) we can check the dependence given by Eq. (\ref{eq:anph}).
This
would corroborate the form of the interaction Hamiltonian chosen
here.

To our knowledge this is the first time that an analysis of
multiabsorption phenomena for massive particles is presented in
the literature. It can be important in situations with highly coherent beams, such as those generated by
atom lasers \cite{AL}, Bose-Einstein condensates \cite{BE}... In the case of
identical bosons in the same state the transition rate of the
double absorption process is proportional to $|\psi |^4$, clearly
different from the usual one for one-particle processes ($\sim
|\psi |^2$). Moreover, the second order transition rate shows very
different behaviours for fermions and bosons. We remark that this
dependence on $|\psi |^4$ provides the basis for a test, in
principle experimentally feasible, of the multiparticle effects
and of the model. With atom lasers the realization of an
experiment which could distinguish between the $|\psi |^2$ and
$|\psi |^4$ behaviours seems to be a realistic goal.

\end{document}